\documentclass[fleqn,usenatbib]{mnras}

\usepackage{newtxtext,newtxmath}

\usepackage[T1]{fontenc}

\DeclareRobustCommand{\VAN}[3]{#2}
\let\VANthebibliography\thebibliography
\def\thebibliography{\DeclareRobustCommand{\VAN}[3]{##3}\VANthebibliography}

\usepackage{graphicx}	
\usepackage{amsmath}	
\usepackage{ulem}

\title[Aspherical ULDM Collapse]{Aspherical ULDM Collapse: Variation in the Core-Halo Mass Relation}

\author[E. Kendall, M. Gosenca and R. Easther]{
Emily Kendall,$^{1}$\thanks{Email: emily.kendall@auckland.ac.nz}
Mateja Gosenca$^{2}$, and
Richard Easther$^{1}$
\\
$^{1}$Department of Physics, University of Auckland, Private Bag 92019, Auckland, New Zealand \\
$^{2}$Faculty of Physics, University of Vienna, Boltzmanngasse 5, 1090 Vienna, Austria \\
}

\date{Accepted XXX. Received YYY; in original form ZZZ}

\pubyear{2023}
\allowdisplaybreaks
\begin{document}
\label{firstpage}
\pagerange{\pageref{firstpage}--\pageref{lastpage}}
\maketitle

\begin{abstract}
Ultralight dark matter (ULDM) is an interesting alternative to the cold dark matter (CDM) paradigm. Due to the extremely low mass of the constituent particle ($\sim10^{-22}$ eV), ULDM can exhibit quantum effects up to kiloparsec scales. In particular, runaway collapse in the centres of ULDM halos is prevented by quantum pressure, providing a possible resolution to the `core-cusp problem' of CDM. However, the the detailed relationship between the ULDM core mass and that of the overall halo is poorly understood. We simulate the collapse of both spherical and aspherical isolated ULDM overdensities using \textsc{AxioNyx}, finding that the central cores of collapsed halos undergo sustained oscillatory behaviour which affects both their peak density and overall morphology. The variability in core morphology increases with the asphericity of the initial overdensity and remnants of initial asphericity persist long after collapse. Furthermore, the peak central densities are higher in spherical configurations. Consequently, astrophysically realistic halos may exhibit substantial departures from theoretical core-halo profiles and we would expect a significant variance of the properties of halos with the same mass. 
\end{abstract}




\section{Introduction}\label{sec:axionyx_intro}

Ultralight (or `Fuzzy') Dark Matter (ULDM) is a bosonic dark matter model in which the constituent particle has an extremely small mass, of order $10^{-22}$eV \citep{Hui:2016ltb}. The corresponding de Broglie wavelength can approach kiloparsec scales and ULDM thus exhibits astrophysically relevant wave-like quantum effects. \citep{Kim:2015yna}. This leads to the suppression of sub-galactic structure relative to Cold Dark Matter (CDM) which has led many to speculate that it may provide a natural resolution to the `small-scale crisis' of CDM \citep{2011MNRAS.415L..40B, Nakama:2017ohe,Robles:2018fur}. 

One aspect of the small-scale crisis, referred to as the `core-cusp problem', describes the tendency of CDM-only simulations to produce dark matter halos with steep or `cuspy' inner density profiles, while observations instead favour smooth `cored' profiles \citep{Bullock:2017xww, DelPopolo:2016emo}. However, the seriousness of the core-cusp problem is the subject of ongoing debate. While some argue that the steepness of observed central density profiles may be underestimated  \citep{2022MNRAS.512.1012C, 2018MNRAS.474.1398G}, others contend that better modelling of baryonic feedback mechanisms \citep{DelPopolo:2015nda} and gas motion in the inner halos may serve to soften central density cusps \citep{2021arXiv211000142J}. These arguments notwithstanding, small-scale discrepancies between CDM-only simulations and observations motivate the study of alternative dark matter models in which these issues may be naturally ameliorated. 

In ULDM, quantum pressure prevents unmitigated gravitational collapse in the centres of halos such that high-density `cusps' do not form. Instead, simulations have shown that while the outer regions of ULDM halos follow Navarro-Frenk-White (NFW) density distributions \citep{Navarro:1995iw}, halo centres are characterised by flattened cores which tend to be well-approximated by the solitonic ground state solution to the Schr\"odinger-Poisson equations \citep{10.1093/mnras/stv1050}. This generic ULDM halo structure has been explored by a number of studies, and an approximate relationship between the mass of the solitonic core ($M_c$) and that of the total virialised halo ($M_h$) has been proposed. Termed the `core-halo' relationship, this was first discussed in \citep{Schive:2014hza}, where it was found $M_c\propto M_h^{1/3}$. Since its inception, however, the applicability of the core-halo relationship has been hotly contested \citep{PhysRevD.94.043513, 10.1093/mnras/stx1887}. Alternative core-halo relationships of the form $M_c\propto M_h^{\beta}$ have been found \citep{PhysRevD.103.063012, 2020ApJ...904..161B}, with the $\beta=1/3$ case seemingly only valid for the most relaxed and spherically symmetric systems \citep{10.1093/mnras/staa3772, Mina:2020eik}. In general, both simulations and analytical investigations point to significant scatter in core characteristics of ULDM halos \citep{Taruya:2022zmt}. 

It is easy to see that such a relationship must be at best approximate, since the cores of ULDM halos are not strictly solitonic but vary over time due to the interference of excited eigenmodes. This interference typically manifests as quasi-normal oscillatory behaviour \citep{PhysRevD.103.023508}. Different physical processes may perturb the core in distinct ways, exciting different eigenmodes and resulting in a core-halo relationship which is dependent upon the  details of halo formation. Indeed, it has been shown that merger history has a non-negligible impact on the core-halo relationship of simulated ULDM halos \citep{PhysRevD.95.043519, Zagorac:2022xic}. Furthermore, it has been shown that stable artificial ULDM halos can be constructed with a wide variety of halo characteristics which need not satisfy the core-halo relationship \citep{Yavetz:2021pbc}.

Variability in the core-halo mass relationship is therefore key to many assessments of the ULDM model  \citep{Kendall:2019fep, Chan:2021bja, PhysRevD.103.063012}. Here we seek to develop insight into core-halo variability from a `first principles' perspective; investigating the collapse of isolated overdensities with tightly controlled initial conditions. We examine the impact of asphericity in the initial overdensity on the behaviour of the core post-collapse. This is motivated by the fact that the peaks of a Gaussian random field, such as those seen cosmological initial conditions, are inherently aspherical \citep{Doroshkevich1970, 1986ApJ...304...15B}. While the generic features of aspherical collapse in pressureless CDM models are well-established \cite{1970A&A.....5...84Z}, a systematic analysis of Zel'dovich-type triaxial collapse in ULDM is lacking. 

Clearly the dynamics of gravitational collapse will differ markedly between the CDM and ULDM models since the existence of quantum pressure in the latter case will eventually cause infalling matter shells to turn around and expand. In turn, this leads to a finite central overdensity which oscillates in time \citep{PhysRevD.99.043540}. In this work we study this phenomenon and its impact on the core-halo relation using numerical simulations. 

The core-halo relation originally derived by \cite{Schive:2014hza} was extracted from a cosmological simulation with Gaussian initial conditions. As such, the initial overdensities will have been naturally aspherical \citep{1986ApJ...304...15B}, however the impact of initial asphericity is not quantified. Moreover, the non-trivial merger history in cosmological simulations means that it may be not be possible to isolate the impact of variability in individual initial overdensities on the final core-halo properties. Finally, the low resolution of individual halos in a cosmological simulation may not fully capture  interference effects. 

To address these limitations, we study the collapse of isolated overdensities with varying initial asphericity in an expanding, matter-dominated, ULDM-only universe using the \textsc{AxioNyx} simulation suite \citep{Schwabe:2020eac}. We find that the shape of the initial overdensity significantly changes the dynamics of collapse. In particular, purely spherical initial overdensities preserve this symmetry through to the post-collapse evolution, introducing what is likely to be an unphysical coherent oscillation in the resulting core. By contrast, biaxial and triaxial initial configurations take longer to collapse than the spherical case. The resulting halos carry imprints of the original asphericity and exhibit more complicated core morphologies. These can deviate significantly from the soliton solution, so that the core-halo mass relation (in which the core is modelled as a soliton) is, at times, ill-defined. 

All of our simulations show significant oscillations in the central density post-collapse, consistent with  previously studies \citep{2021ApJ...916...27D, Chiang:2021uvt, Schwabe:2021jne} 
These oscillations persist without significant damping over cosmological timescales resulting in strong time dependence in the instantaneous core-halo mass relationship and the shape of the core itself.

This work is organised as follows. In Section \ref{sec:uldm} we introduce our notation and the equations describing the underlying dynamical system. We also give a brief review of the \textsc{AxioNyx} simulation suite. In Section \ref{sec:biaxial} we present results for a `reference' spherical collapse and biaxial initial states, while we consider triaxial scenarios in Section \ref{sec:triaxial}. We conclude in Section \ref{sec:axionyx-concl} and discuss avenues for further research.

\section{Ultralight Dark Matter Dynamics}\label{sec:uldm}

At late times in an expanding universe ULDM is modelled as a non-relativistic coherent scalar field, $\psi$, governed by the comoving Schr{\"o}dinger-Poisson equations. 
Denoting the scale factor by $a$ we have:
\begin{align}\label{eq:com_sh}
    &i\hbar \, \dot{\psi} = -\frac{\hbar^2}{2ma^2}\nabla^2\psi + m\Phi\psi,\\
    \label{eq:com_p}
    &\nabla^2\Phi = \frac{4\pi G m}{a} \vert\psi\vert^2.
\end{align}
where $G$ is the gravitational constant, $\Phi$ is the gravitational potential, and $m$ is the mass of the ULDM particle. The physical density is $\rho = m \vert \psi\vert^2 / a^3$. These equations are numerically integrated to simulate the evolution of self-gravitating ULDM in the absence of other interactions.  

Previous simulations  suggest that relaxed ULDM halos generically consist of a smooth central core surrounded by an incoherent outer halo \citep{Veltmaat:2018dfz, PhysRevD.94.043513, Lin:2018whl}. The spherically-averaged profile of the outer halo has been found to closely match the Navarro-Frenk-White (NFW) profile of CDM \citep{Navarro:1995iw}:
\begin{equation}
    \rho_{\text{NFW}}(r) = \frac{\rho_0}{\frac{r}{R_s}\left(1+\frac{r}{R_s}\right)^2},
\end{equation}
where $\rho_0$ and $R_s$ are free parameters which vary from halo to halo.  Meanwhile, the inner core tends to resemble the solitonic ground-state of the Schr{\"o}dinger-Poisson equations. Its physical density profile is well-approximated by \citep{Schive:2014dra}
\begin{equation}\label{eq:dens_prof}
    \rho_c(r) = \frac{1.9\times 10^7 m_{22}^{-2}(r_c/\mathrm{kpc})^{-4}}{\left(1+0.091(r/r_c)^2\right)^8}\mathrm{M}_{\odot}\mathrm{kpc}^{-3}
\end{equation}
where $m_{22} = m/10^{-22}\text{eV}$ and $r_c$ is the physical core radius, defined heuristically by the density dropping to half-maximum. The core radius is inversely proportional to the core mass, $M_c$, and \cite{Schive:2014hza} find that it is related to the total halo mass by
\begin{equation}\label{eq:core-halo_alt}
    r_c = 1.6 \, m_{22}^{-1} a^{1/2}\left({M_h}/{10^9 M_\odot}\right)^{-\beta}\text{kpc},
\end{equation}
where $M_h = (4\pi r_\text{vir}^3/3)\zeta(z)\rho_\text{crit}$ is the virial mass of the halo and $\beta = 1/3$ in \cite{Schive:2014hza}. For simplicity, in this work we assume a matter dominated universe with $\zeta(z) = 200$ for all $z$ \citep{Bryan:1997dn}. The scale factor, $a$, appears in Equation \ref{eq:core-halo_alt}, reflecting the empirical finding that the core-halo  relationship is redshift-dependent. While Equation \ref{eq:core-halo_alt} was derived from the results of a cosmological simulation, higher resolution simulations of simplified scenarios such as individual halo collisions are also suggestive of a broadly applicable core-halo relationship. However, there is some argument as to whether $\beta={5/9}$ yields a better fit than $\beta={1/3}$ \citep{Mina:2020eik,10.1093/mnras/stx1887} and recent work indicates that the best-fit scaling is sensitive to halo formation history \citep{Zagorac:2022xic}.

\begin{figure*}
\centering
\includegraphics[scale = 0.54, trim={0.5cm 0.5cm 0cm 0cm}]{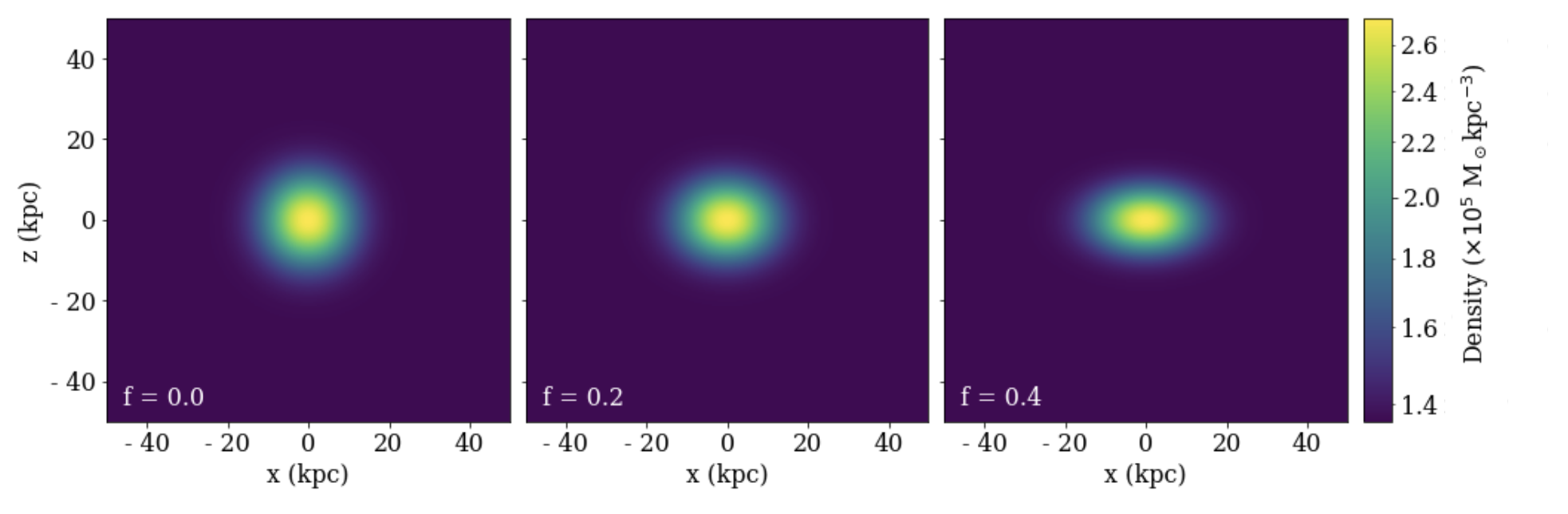}
\caption{Initial field configurations ($z=9$) in the biaxial collapse scenarios for different values of the flattening parameter. As $f$ is increased, the overdensity is flattened along the $z-$direction (vertical axis). The physical box side length is 100 kpc.}\label{fig:initial_config}
\end{figure*}

We initialise our simulations with isolated overdensities that are Gaussian functions of $r_\text{ell}$:
\begin{equation}
    r_\text{ell} = \sqrt{\left(\frac{x}{A}\right)^2 + \left(\frac{y}{B}\right)^2 + \left(\frac{z}{C}\right)^2},
\end{equation}
where $A$, $B$, and $C$ are varied to create aspherical configurations. The initial density profile at redshift $z_\text{init}$ is
\begin{equation}\label{eq:symm_case}
    \rho(x,y,z) = \rho_{\text{crit}}(z_\text{init})\left(1 +\delta\exp\left(-\frac{r_\text{ell}^2}{r_\text{ch}^2}\right)\right),
\end{equation}
where $r_\text{ch}$ is a characteristic length scale which determines the width of the overdensity and the density approaches the critical density, $\rho_{\text{crit}}$, at large distances from the centre of the overdensity. The initial velocity of the matter is set to zero.  

All simulations were run in a box with comoving side length $L = 1.0 \text{ Mpc}$. Our parameter choices are $z_\text{init}=9$ ($a_\text{init}=0.1$), $r_{\text{ch}} = 0.1 a_{\text{init}} L$, and $\delta = 1.0$. This choice of initial redshift may appear low, however we find that in all cases the redshift at collapse and the post-collapse behaviour are very similar if the simulations are instead initialised at $z_\text{init}=99$ with $\delta = 0.1$. Given that either choice of initial conditions is somewhat artificial, we opt for the later start as this allows for a higher spatial resolution in the ``present'' epoch at the same level of grid refinement. The width $r_\text{ch}$ has been chosen to ensure the collapse of the overdensity while limiting the impact of periodic boundary conditions. We will see that these parameter choices correspond to final ULDM halos with virial masses of order $10^9\text{M}_\odot$.

We consider both biaxial (flattened) and triaxial overdensities. In all cases we keep the product $ABC = 1$. Integrating the second term of Equation \ref{eq:symm_case} shows that under this condition the mass of the initial overdensity is independent of its shape and equal to 
\begin{equation}
    M = \delta \pi^{3/2} \rho_\text{crit}(z_{\text{init}}) r_{\text{ch}}^3.
\end{equation}

We construct biaxial (flattened) overdensities, by setting $A = B > C$ and characterise the flattening via 
\begin{equation}
    f = \left(A-C\right)A^{-1}.
\end{equation}
We vary $f$ between  $0.0 \leq f \leq 0.5$ while keeping the  $ABC \equiv 1$ so that
\begin{equation}
    A = B = \left({1-f}\right)^{-1/3}, \quad \text{and} \quad C = (1-f)^{2/3} .
\end{equation}
The restriction to positive values of $f$ is consistent with  expectations for generic Gaussian random fields \citep{1986ApJ...304...15B}, which suggest that overdense regions tend to be oblate.  

For triaxial configurations we have $A > B > C$ and define the ratio $r$ such that:
\begin{equation}\label{eq:triax_config}
    A = 1/r, \quad B = 1, \quad C = r.
\end{equation}
Consequently the product $ABC = 1$ and the ratios $A/B$ and $B/C$ are equal to one another. This restriction is obviously a departure from full generality but permits a useful initial survey of the influence of asphericity on ULDM halo formation. 

Our simulations are performed with \textsc{AxioNyx} \citep{Schwabe:2020eac}.  This code is built on top of the \textsc{Nyx} cosmology simulator \citep{Almgren:2013sz}, which is itself built upon  \textsc{AMReX}  \citep{amrex2021}, a sophisticated adaptive mesh refinement (AMR) library. \textsc{AxioNyx} implements both pseudospectral and finite-differencing methods to solve the Schr\"{o}dinger-Poisson system, using the former on the root grid, and the latter in refined regions where periodic boundary conditions no longer apply. 

All simulations are initialised on a $512^3$ base grid with up to 5 levels of refinement. With $H_0 = 70 \text{km} \text{s}^{-1} \text{Mpc}^{-1}$  the critical density is $\rho_\text{crit}(a) = a^{-3}\times 1.36\times 10 ^{11}\text{ M}_\odot\text{Mpc}^{-3}$ and we set $m_{22} = 1$ throughout. Some care is needed when specifying the refinement criteria in ULDM simulations. The outer regions of a halo can contain low-density, high-velocity material resulting from the expulsion of matter in shock waves.  In naive schemes this may trigger unnecessary refinement at  massive computational cost as these regions can account for a substantial fraction of the overall simulation volume. To focus on the dynamics of the central halo regions we implement a cascade of increasing density thresholds rather than a single velocity-dominated threshold to trigger refinement.

\section{Spherical and Biaxial Collapse} \label{sec:biaxial}

\begin{figure*}
\centering
\includegraphics[scale=0.5, trim={0cm 0cm 0cm 0cm}]{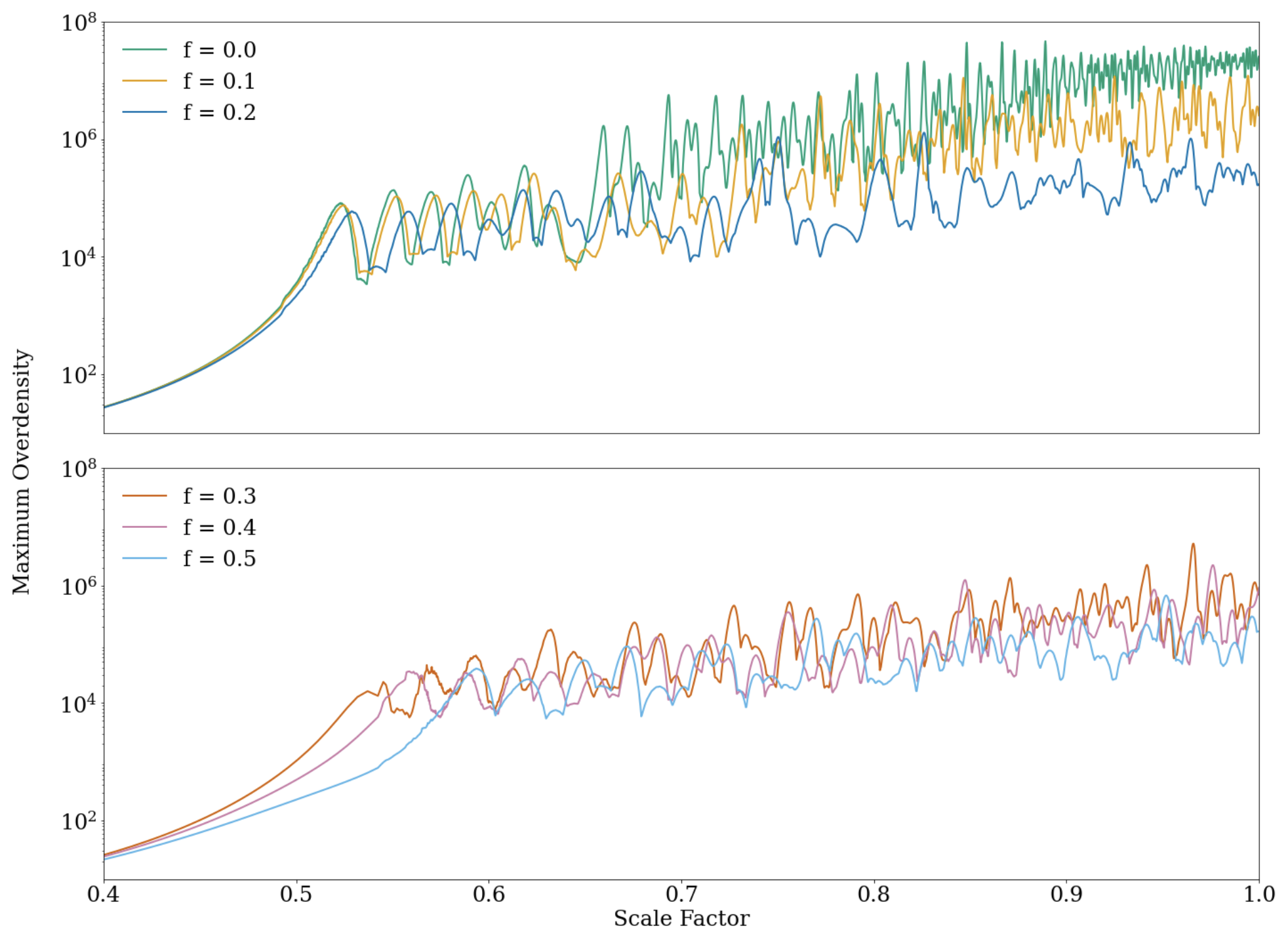}
\caption{The maximum value of the dimensionless overdensity, $(\rho -\bar{\rho})/\bar{\rho}$, as a function of the scale factor for each of the biaxial collapse simulations. Collapse time is indicated by the first peak in the overdensity curve, which occurs later for larger $f$. The initial overdensities have the same mass but the maximum overdensity after collapse is much larger for approximately spherical configurations, indicating that core-halo characteristics are sensitive to initial shape. Note that the vertical scale is the same in both plots, which we have separated for clarity.} 
\label{fig:max_overdens_bi}
\end{figure*}

We begin with biaxial overdensities with flattening parameters,  $f \in \{0.0, 0.1, 0.2, 0.3, 0.4, 0.5\}$. Figure \ref{fig:initial_config} illustrates the  shape of the initial overdensities, where the $f=0$ case is the sperical limit. 

In Figure \ref{fig:max_overdens_bi} we illustrate the collapse process by plotting the maximum overdensity within the simulation box as a function of the scale factor. As expected, the overdensity initially grows under its own self-gravity but eventually reaches a peak and rebounds as a result of the repulsive ULDM quantum pressure. We define the moment of `collapse' as the instant at which the   overdensity reaches its first maximum. Collapse is followed by a sustained oscillatory phase, as  previously described in \citep{PhysRevD.101.083518,Schwabe:2020eac}. This oscillatory behaviour is due to interference between the ground and excited states of the Schr\"odinger-Poisson system. These oscillations can be large and highlight the extent to which ULDM halo cores may not always be well-modelled as stable (ground state) solitons. 

The limiting case of $f=0.0$ represents purely spherical collapse, which occurs shortly after $z=1$ ($a=0.5$) in these simulations. However, as $f$ is increased, collapse time also increases, as shown in Figure \ref{fig:max_overdens_bi}. The initial overdensities all have the same mass, so this effect illustrates the dependence of collapse history on overdensity shape alone. It is consistent with the understanding that isolated ellipsoidal overdensities collapse most slowly along their longest axes \citep{Mead:2016ybv, 1970A&A.....5...84Z} and is not unique to ULDM \citep{Okoli:2015dta, Sheth:1999su}. In more complicated structure formation scenarios, external tidal shears would be expected to have a non-trivial effect on collapse history as well \citep{1995ApJ...439..520E}.


Another notable feature revealed by Figure \ref{fig:max_overdens_bi} is that the peak  overdensities tend to decrease as $f$  increases. Moreover, the height of the typical maximum overdensity is still increasing at the end of the simulation for the  spherical case while it has approximately plateaued for  higher values of $f$. Indeed, at $z=0$ the maximum overdensity for the purely spherical case is several orders of magnitude higher than for $f\geq0.2$. While limited spatial resolution may affect predictions at late times, the correlation between asphericity and the maximum overdensity is clearly seen early in the oscillatory phase where the de Broglie wavelength is comfortably resolved.

Given that the overdensities have the same initial masses, it is reasonable to expect that the virial masses of the resulting collapsed halos will be similar. We verify this numerically, finding that at $z=0$ the virial masses differ by at most a factor of 1.3 while this disparity decreases at earlier times. Given the similarity in the virial masses of the collapsed halos at any given time, the core-halo relation in Equation \ref{eq:core-halo_alt} predicts very similar core density profiles. By contrast, Figure \ref{fig:max_overdens_bi} illustrates that the central densities of the collapsed halos tend to decrease as $f$ is increased. Because $\rho_c\propto M_c^4$ for the theoretical soliton profile, this indicates a substantial variability in core mass and thus a single core-halo relation cannot fit all curves simultaneously. Indeed, a single relation cannot fit even one such curve given the strong oscillations in the maximum density over time.

\begin{figure*}
\centering
\includegraphics[scale=0.49, trim={0cm 0cm 1cm 0cm}]{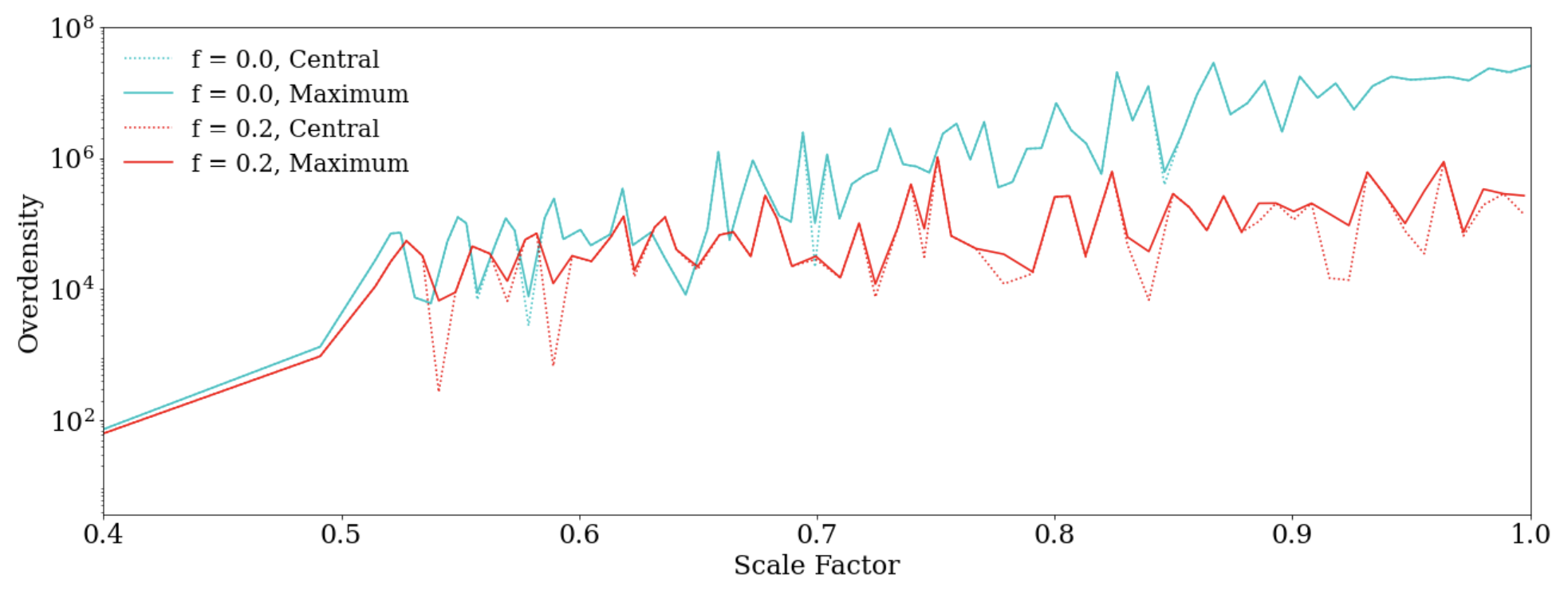}
\caption{Comparison of the maximum overdensity and the central overdensity for $f=0.0$ and $f=0.2$. The maximum overdensity not being centrally located is indicative of a non-solitonic core morphology.}
\label{fig:max_v_cent}
\end{figure*}

A final observation is that the maximum overdensity curves occasionally exhibit discontinuities in their first derivatives. Physically, this occurs because the point of maximum overdensity is not always located at the centre of the simulation region. Therefore, while the overdensity at any fixed point is a smoothly varying function of time, the maximum overdensity curve exhibits abrupt changes when location of the global maximum shifts. We plot both the central overdensity and the maximum overdensity  for $f=0.0$ and $f=0.2$ in Figure \ref{fig:max_v_cent} to illustrate this effect.\footnote{Note that while the maximum overdensity is saved at every simulation timestep, the full field configuration (and hence the central overdensity) is saved less frequently. Consequently, the temporal resolution in these plots is decreased relative to Figure \ref{fig:max_overdens_bi} and they exhibit sharper features.} The discrepancy between the central and maximum overdensity curves becomes more pronounced with greater asphericity, corresponding to cores which deviate more strongly from the spherically symmetric solitonic profile. Interestingly, the maximum overdensity and central overdensity curves in Figure \ref{fig:max_v_cent} tend to align at their local maxima while deviations tend to occur at local minima. This indicates that the core periodically returns to an approximately solitonic profile but assumes a more complicated morphology throughout the remainder of the oscillation. This is consistent with previous observations of complicated core behaviours, which may be analysed using an eigenstate decomposition in the limit that the potential becomes quasi-static \citep{Zagorac:2021qxq}.

\begin{figure*}
\centering
\includegraphics[scale=0.52, trim={6cm 2cm 0cm 2cm}]{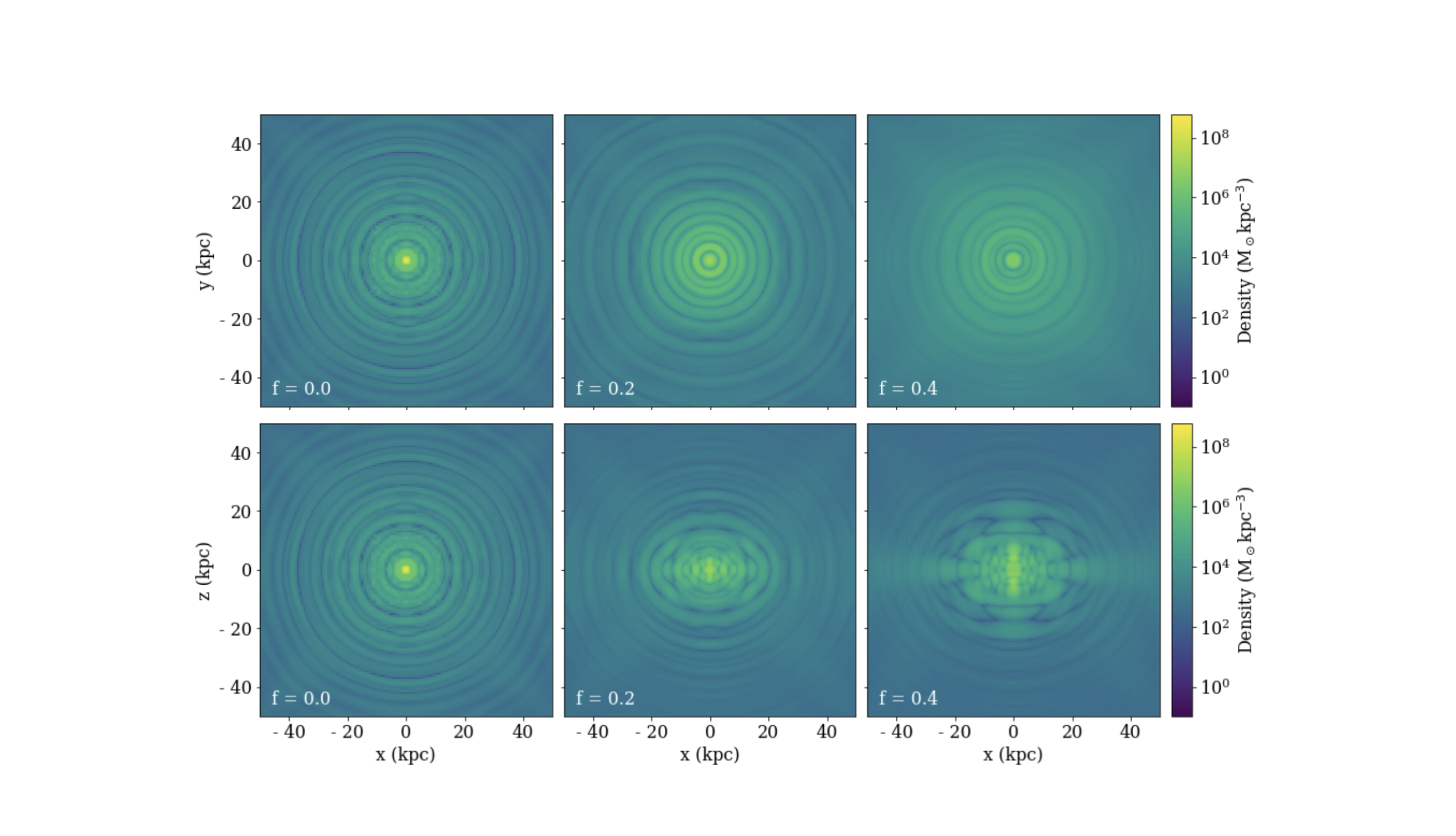}
\caption{Post-collapse density distributions across the $x-y$ (top) and $x-z$ (bottom) planes for $f=0.0$, $f=0.2$, and $f=0.4$. These snapshots are taken at redshift 0.3, which is well into the oscillatory phase in all cases. For clarity, only the innermost (physical) 100kpc of a much larger simulation region are shown. Because the initial overdensities are only flattened along the $z$ direction, the distributions appear circular in the $x-y$ plane with coherent concentric wavefronts. By contrast, the wavefronts undergo non-trivial interference in the $x-z$ plane, yielding more complicated density distributions.} 
\label{fig:biax_slice}
\end{figure*}

Figure \ref{fig:biax_slice} shows snapshots of the density profiles through the centres of collapsed halos at $z=0.3$ for $f=0.0$, $f=0.2$ and $f=0.4$.\footnote{We choose $z=0.3$ as a reference case since at this redshift all simulations are well within the post-collapse oscillatory phase, while the finest structures are still well resolved. Toward the end of the simulations, particularly in the $f=0$ case, the spatial resolution may not be high enough to accurately model the behaviour in the core.}  As a consequence of our idealised initial conditions we do not see the random `granular' outer regions typically associated with ULDM halos \citep{Schwabe:2021jne, Veltmaat:2018dfz}. The initial symmetry is well preserved by \textsc{AxioNyx} and our simulations instead exhibit coherent wave fronts propagating outward from the centre. However, as the flattening parameter is increased, these wave fronts become distorted and the interference effects become more complicated. This leads to more complicated density distributions, and for $f=0.4$ we see evidence of an approximately granular structure reminiscent of `typical' ULDM halos. We conclude that in order for truly random incoherent outer halos to develop, one or more explicit symmetry-breaking mechanisms are required. This could be achieved through mergers, interactions with randomly distributed neighbouring overdensities or additional (e.g. baryonic) matter components inside the halo.

\begin{figure}
\centering
\includegraphics[scale=0.25, trim={0cm 0cm 0cm 0cm}]{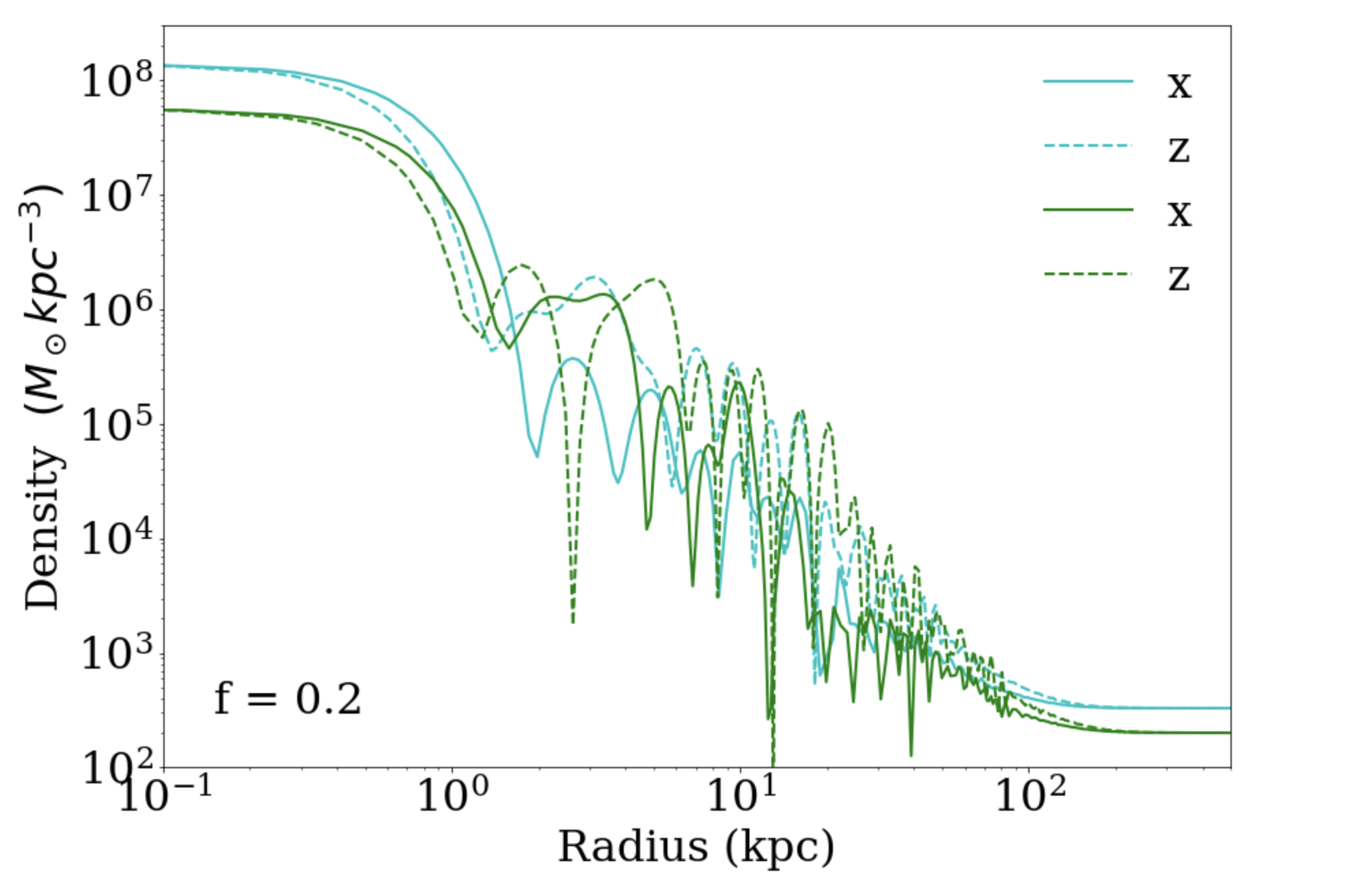}
\caption{Comparison of the density profiles along the $x$ and $z$ axes in the $f=0.2$ biaxial collapse simulation at redshift $z=0.352$ (blue) and $z=0.148$ (green). While the cores appear solitonic under spherical averaging, the profiles retain imprints of the initial flattening.} 
\label{fig:x_vs_z}
\end{figure}

Throughout the post-collapse oscillatory phase, the cores frequently become significantly distorted relative to the solitonic profile described by Equation \ref{eq:dens_prof}. Indeed, it appears that the solitonic profile only provides a reasonable fit to the cores at or near the local maxima of the overdensity curves of Figure \ref{fig:max_overdens_bi}. Furthermore, it is important to note that even when the cores appear roughly solitonic under spherical averaging, asphericity may still persist. In Figure \ref{fig:x_vs_z} we show two representative density profiles drawn from the $f=0.2$ simulation at times when the core appears roughly solitonic. In both cases, the core profile is systematically wider along the $x-$axis and narrower along the $z-$axis, echoing the flattening in the initial overdensity. 

The profiles shown in Fig.~\ref{fig:x_vs_z} were deliberatley chosen at instances when the cores are roughly solitonic under spherical averaging. However, the strong oscillations in the core change the $x$ and $z$ profiles over time, leading to highly variable core masses and morphologies. This suggests that the piece-wise `solitonic core + NFW outer halo' ULDM halos template \citep{Robles:2018fur} is not universally valid, and reduces the generality of the theoretical core-halo mass relationship \citep{Schive:2014hza, Niemeyer:2019aqm}. These results are consistent with several recent studies. For example, \cite{PhysRevD.103.023508} find that perturbations of a ULDM halo core lead to  significant variability in the mass of the inner halo. This is likely to have consequences for observational tests of ULDM given the wide diversity of observed halo profiles \citep{Oman:2015xda, 10.1093/mnras/stw1876, Kendall:2019fep}.

\begin{figure}
\centering
\includegraphics[scale=0.25, trim={0cm 0cm 0cm 0cm}]{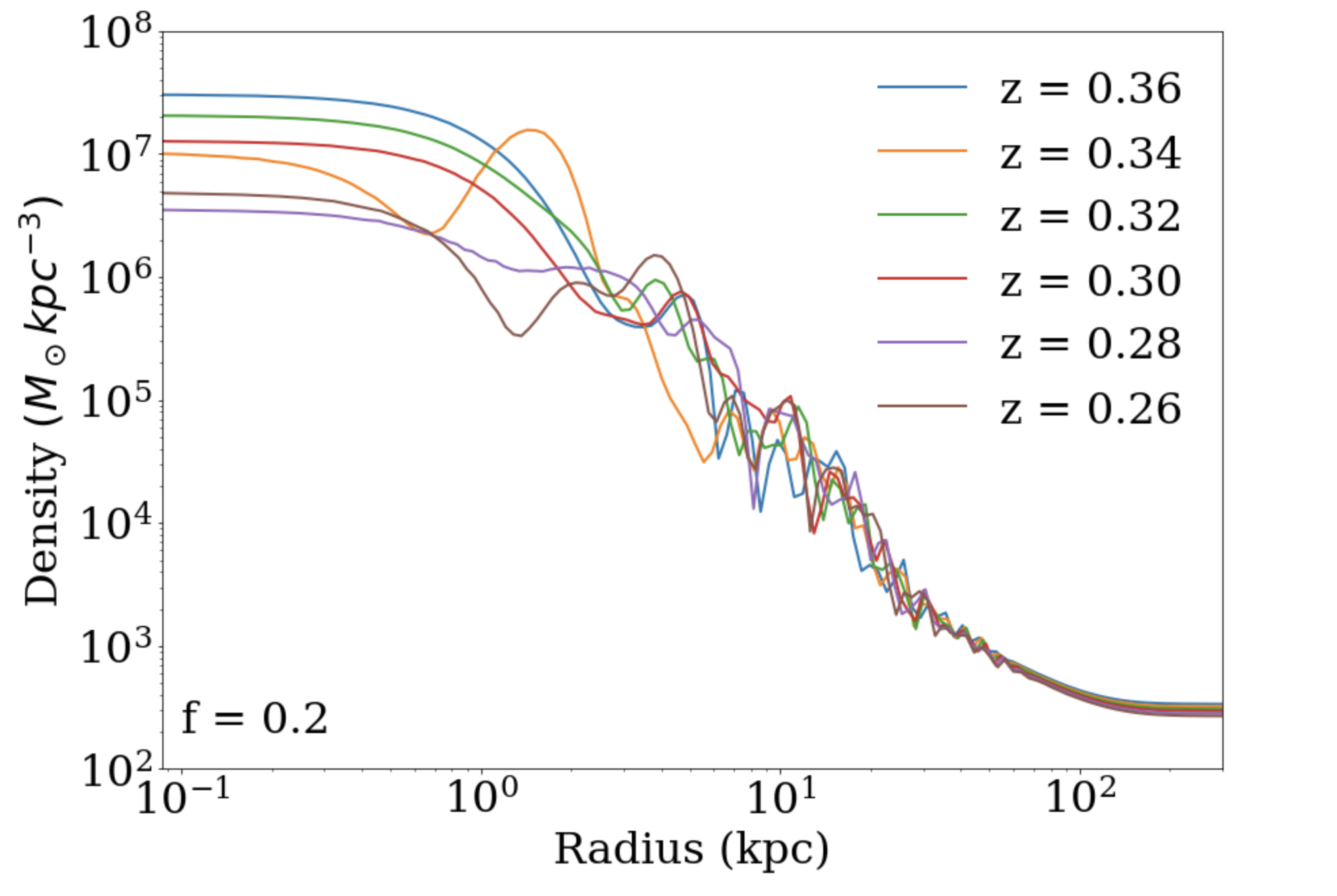}
\caption{Time series of spherically-averaged density profiles during the post-collapse oscillatory phase of the $f=0.2$ biaxial collapse simulation. A wide diversity of core characteristics is seen over a relatively small redshift range.} 
\label{fig:f02_peak_to_peak}
\end{figure}

To explore the diversity of core characteristics we compute the spherically-averaged halo profile over a full oscillation of the maximum overdensity curve for a representative period in the evolution of the  $f=0.2$ halo.  The results of this process are shown in Figure \ref{fig:f02_peak_to_peak}. We see that within a small redshift range the central density varies by an order of magnitude. Furthermore, the morphology of the core (e.g. at $z=0.34$) can differ significantly from a solitonic profile due to interference effects. 

\begin{figure}
\centering
\includegraphics[scale=0.25, trim={0cm 0cm 0cm 0cm}]{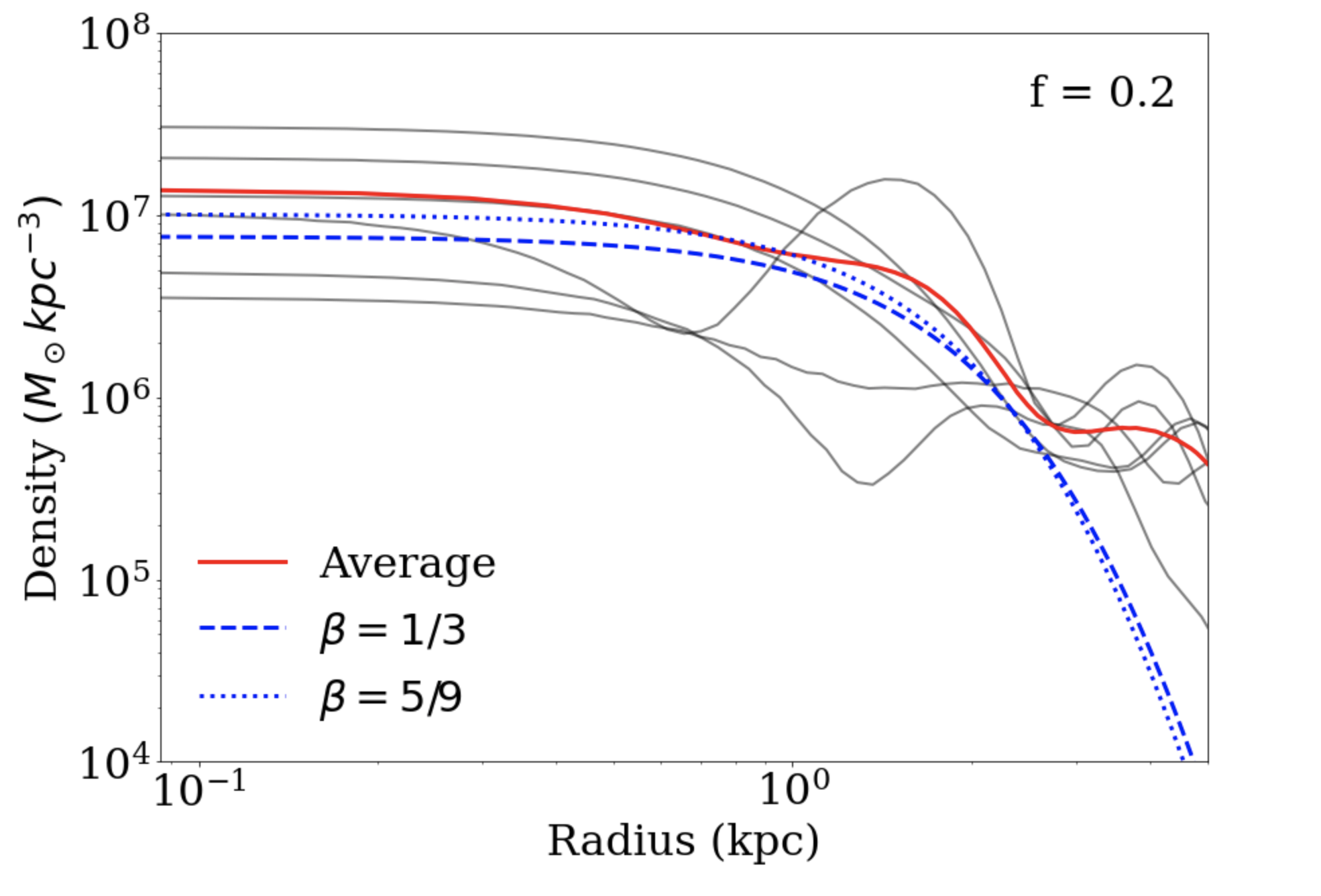}
\caption{Comparison of the innermost five kiloparsec of the halos from Figure \ref{fig:f02_peak_to_peak} and the core profiles predicted by Equation \ref{eq:core-halo_alt} for $\beta=1/3$ (blue dashed line) and $\beta=5/9$ (blue dotted line). Also plotted is an average of the six individual profiles (red line).} 
\label{fig:f02_core_halo}
\end{figure}

To test the validity of the core-halo relation over time  we use Equations \ref{eq:dens_prof} and \ref{eq:core-halo_alt} to produce solitonic core profiles for each snapshot for  $\beta=1/3$ and $\beta=5/9$, given that arguments have been made for both values \citep{Mina:2020eik, 10.1093/mnras/stx1887}, although recent work has suggested that the best-fit scaling is sensitive to formation history \citep{Zagorac:2022xic}. The core oscillation is much faster than the timescale of the redshift-dependence of the core-halo relation or the virial mass\footnote{We compute the virial mass of the collapsed halo for each profile shown in Figure~\ref{fig:f02_peak_to_peak} using the average internal density criterion $\bar{\rho} = 200 \rho_\text{crit}$. Other conventions exist for defining the virial mass \citep{2001A&A...367...27W}, but this choice should have a sub-leading affect on any core-halo analysis.}  so we can treat these values as constant. The results of this process  are shown in Fig.~\ref{fig:f02_core_halo}. For each value of $\beta$ the predicted core profile changes minimally and is plotted as a single line. Neither of the predicted cores provide a good fit to the majority of individual profiles. When the six individual profiles are averaged the result is closer to the theoretical prediction, with $\beta=5/9$ providing the better fit. However, a naive comparison of the theoretical core-halo relation with a time-averaged profile may be misleading, as it does not take into account the complicated relationship between the dark matter density profile and the motion of observable tracer objects. In particular, it has been shown that oscillations in ULDM cores can have a dramatic effect on stellar heating in the inner halo \citep{Dalal:2022rmp}. Detailed simulations including baryons will be undertaken in future work to determine the extent to which strong oscillations in the core affect the radial distribution of tracer objects and their velocity dispersion.

Finally, we observe that while the density profiles in Figure \ref{fig:f02_core_halo} vary by approximately an order of magnitude at their centres, the core mass itself does not change as dramatically. In particular, Equation \ref{eq:dens_prof} shows that $\rho_c(r=0)\propto r_c^{-4}$ for a soliton, and since $M_c\propto r_c^{-1}$ \citep{Edwards:2018ccc} an order of magnitude change in central density corresponds to a change in core mass of less than a factor of 2. Thus the core-halo relation does not seem to be the most informative metric with which to probe the ULDM model, as halos which have similar core masses may have dramatically different inner density profiles. This is true even in the `best case scenario' when the cores are approximately solitonic, and is worsened when the morphology of the cores changes during the oscillations.

\section{Triaxial Collapse}
\label{sec:triaxial}

\begin{figure*}
\centering
\includegraphics[scale=0.5, trim={0cm 0cm 0cm 0cm}]{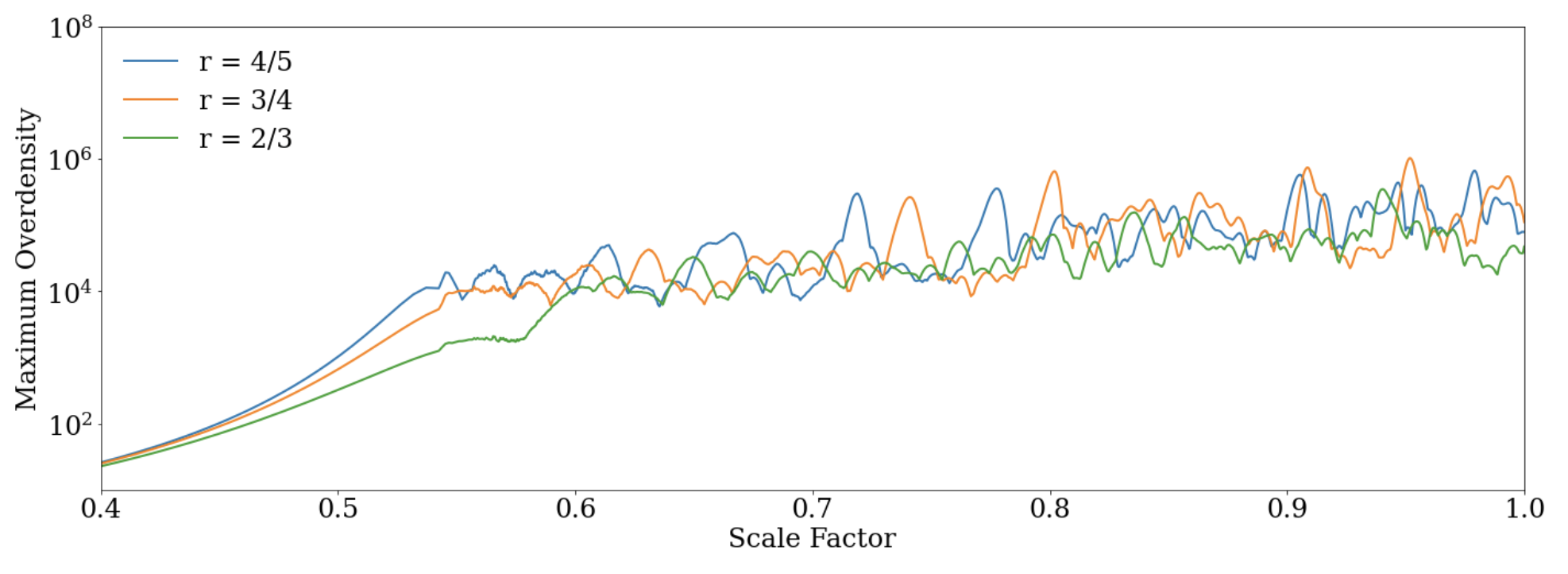}
\caption{Maximum overdensity as a function of the scale factor for each of the triaxial collapse simulations. Collapse time again increases with asphericity, while maximum post-collapse overdensity increases as spherical symmetry is approached.} 
\label{fig:max_overdens_tri}
\end{figure*}

Gaussian random fields such as those describing density fluctuations in the early universe exhibit peaks which are inherently triaxial \citep{10.1093/mnras/stu2021}. We therefore extend our analysis to triaxial configurations, but for simplicity we restrict ourselves to a small subset of possible triaxial configurations, namely those described by Equations \ref{eq:symm_case} and \ref{eq:triax_config}. We evolve initial configurations with  $r\in\{4/5,\, 3/4, \, 2/3\}$, corresponding to increasingly aspherical initial conditions. The maximum overdensity as a function of the scale factor is shown in Figure \ref{fig:max_overdens_tri}. Again, collapse occurs more slowly as asphericity is increased. The maximum overdensity tends to be lowest in the most aspherical case ($r=2/3$), but this suppression is less pronounced than in the biaxial case. 

Significant oscillations in the maximum overdensity for each halo (illustrated in Figure \ref{fig:max_overdens_tri}) again indicate that a single core-halo relation will not be universally applicable. The oscillatory phase appears to be more stochastic for triaxial collapse, which is consistent with more complicated interference effects in this regime. Consequently we expect significant deviations from solitonic cores throughout the oscillatory phase. 

\begin{figure*}
\centering
\includegraphics[scale=0.52, trim={0.5cm 0.5cm 0cm 0cm}]{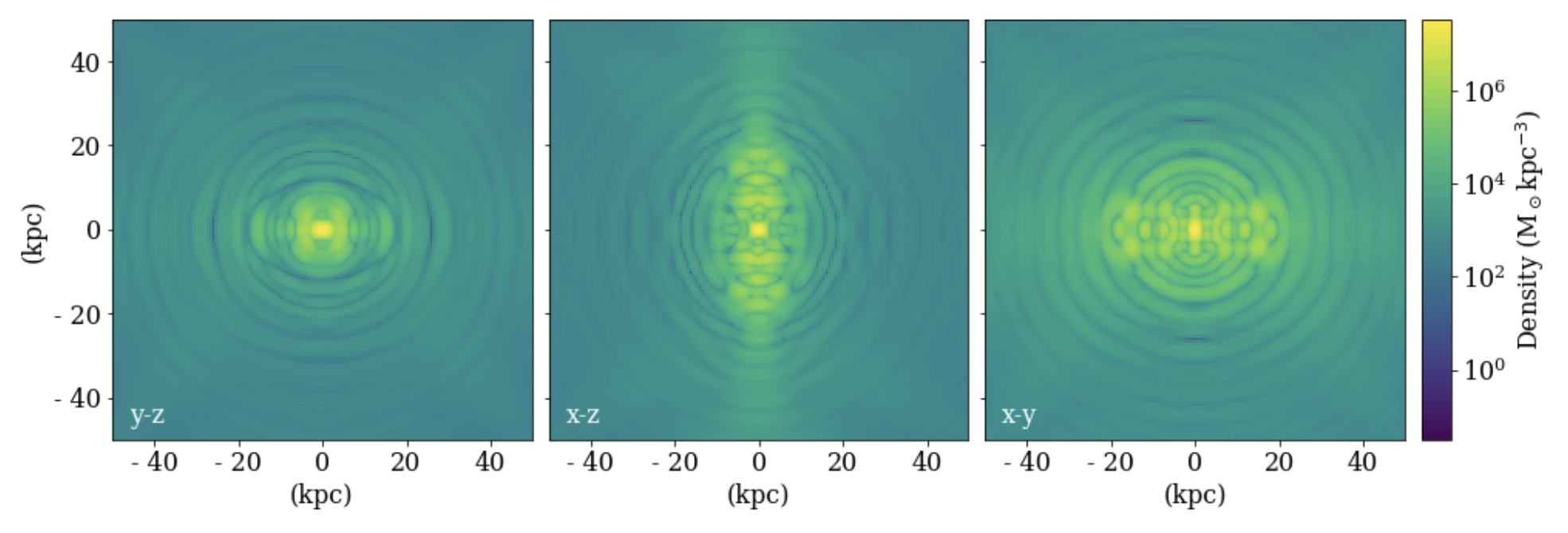}
\caption{Density distributions across each plane (y-z, x-z, x-y) for the $r=4/5$ triaxial collapse simulation at redshift 0.3. Again only the innermost (physical) 100kpc of a much larger simulation region is shown. By contrast with the biaxial collapse scenarios at the same redshift, these distributions exhibit more granularity.} 
\label{fig:dens_slice_tri}
\end{figure*}

Figure \ref{fig:dens_slice_tri} illustrates the central $100$~kpc of the post-collapse halo at $z=0.3$ with $r=4/5$ case. The triaxial interference patterns are more complicated than their biaxial analogues at similar redshift, and the density distribution more closely approaches the granular structure of typical ULDM halos. The central core, which is shown along each of the three axes in Figure \ref{fig:dens_slice_tri}, does not have a spherical solitonic shape but instead we see the development of `lobes' which are particularly evident in the y-z plane. This particular snapshot corresponds to a local minimum in the overdensity for which the maximum density is not centrally located, a time when the core-halo relation is unlikely to provide a good fit. 

\begin{figure}
\centering
\includegraphics[scale=0.25, trim={0cm 0cm 0cm 0cm}]{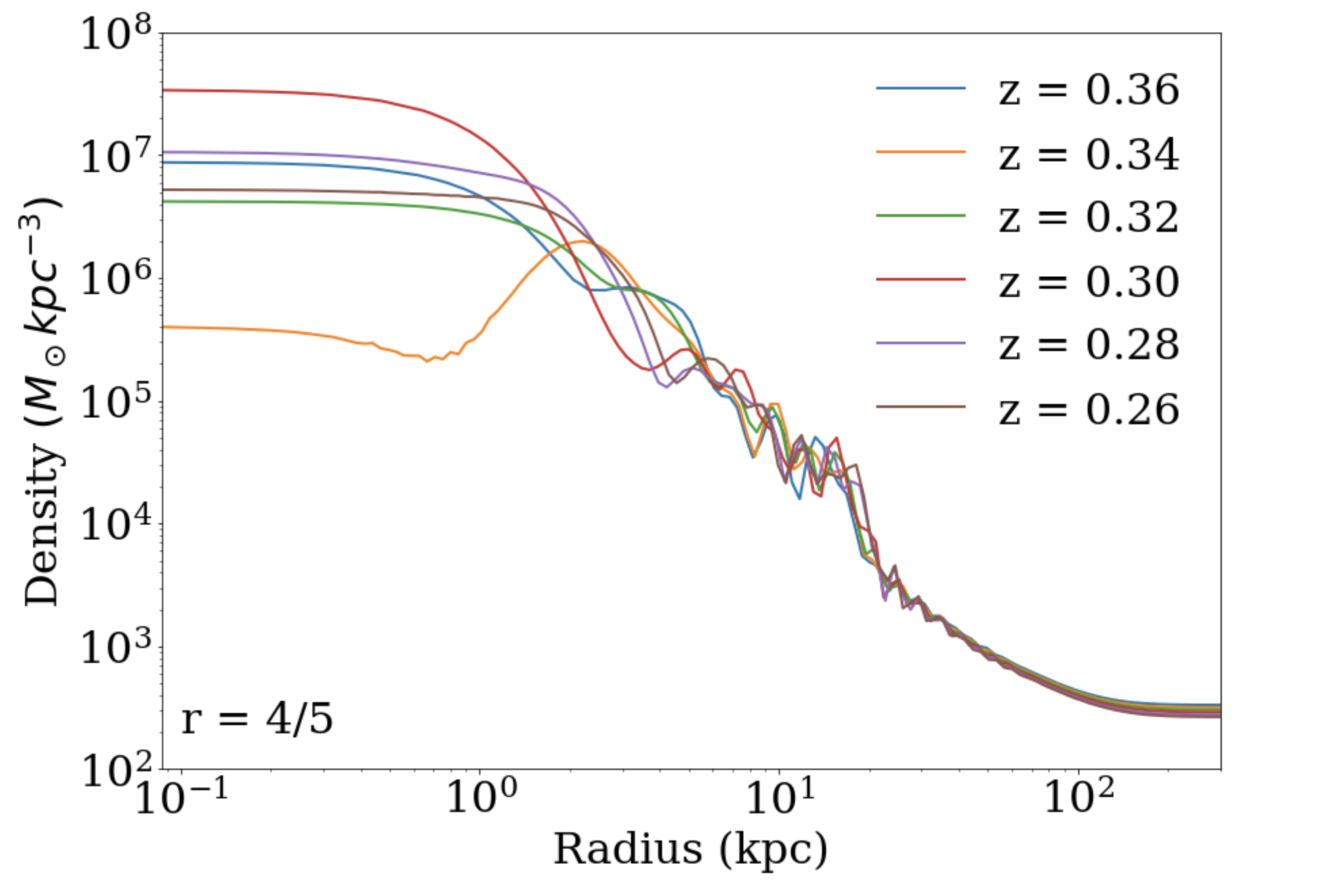}
\caption{Time series of spherically-averaged density profiles during the post-collapse oscillatory phase of the $r=4/5$ triaxial collapse simulation. A wide diversity of core characteristics is again present, with variation in the central density of more than an order of magnitude.} 
\label{fig:r45_peak_to_peak}
\end{figure}

We repeat the previous core-halo analysis for the triaxial $r=4/5$ case. For consistency we  use the same redshift range as in the biaxial $f=0.2$ case ($0.26\leq z\leq 0.36$), although results are similar for other intervals.  We plot the full spherically-averaged density profile in Figure \ref{fig:r45_peak_to_peak} for each snapshot.  Again we see a large diversity in core characteristics, with distortions that can amount to significant deviations from the spherical solitonic profile. 

\begin{figure}
\centering
\includegraphics[scale=0.25, trim={0cm 0cm 0cm 0cm}]{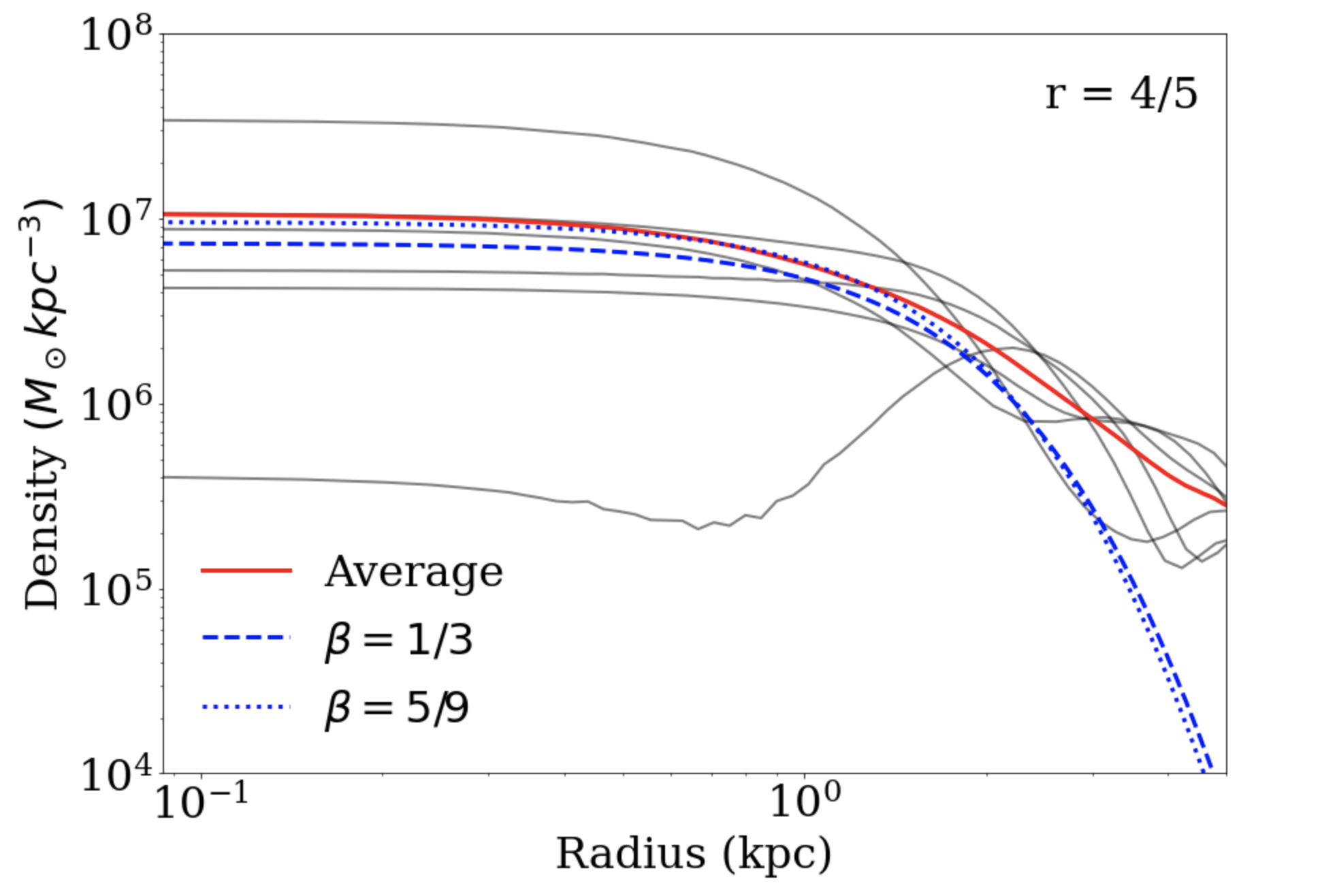}
\caption{Comparison of the innermost five kiloparsec of the halos from Figure \ref{fig:r45_peak_to_peak} and the core profiles predicted by Equation \ref{eq:core-halo_alt} for $\beta=1/3$ (blue dashed line) and $\beta=5/9$ (blue dotted line). Also plotted is an average of the six individual profiles (red line).} 
\label{fig:r45_core_halo}
\end{figure}

For each snapshot shown in Figure \ref{fig:r45_peak_to_peak} we compute the virial masses and predicted core profiles, noting that both remain approximately constant throughout this limited redshift range. We compare the individual and averaged profiles to the predicted cores (for both $\beta = 1/3$ and $\beta = 5/9$) in Figure \ref{fig:r45_core_halo}. Again, the diversity of density profiles preclude good individual fits to either of the predicted cores, but the average profile is well fit by $\beta=5/9$. However, because the factor $M_h/10^9M_\odot\sim 1$ for all of our simulations it is difficult to test the $\beta$ dependence directly. While our results tentatively suggest $\beta=5/9$ provides a better fit to simulation data, future simulations over a broader mass range will be required to confirm this. 

\section{Discussion}\label{sec:axionyx-concl}

In this work we simulated the collapse of isolated overdensities with varying degrees of initial asphericity, including both biaxial and triaxial configurations in an expanding ULDM universe. This analysis is an exploratory survey of the impact of initial conditions on the shape and evolution of ULDM halos and raises a number of questions for future investigation. 

All collapse simulations show strong oscillations in the cores of the collapsed halos which persist over cosmological timescales. This is consistent with previous simulations and is a consequence of the interference of excited field modes \citep{PhysRevLett.124.201301, Zagorac:2021qxq, PhysRevD.69.124033, PhysRevD.101.083518}. We find that the core oscillations in the spherically symmetric case are much stronger than for halos with more complicated geometries. However, such a highly symmetric configuration is of course unphysical, and oscillations in realistic halos are likely to be suppressed relative to this extreme case. 

We find that the collapse timescale increases with the initial asphericity, which is consistent with the principle that overdensities collapse most slowly along their longest axes. In the post-collapse phase, halos preserve remnants of initial asphericity. This is true both in the outer halo and in the core, which may deviate from the spherically symmetric solitonic profile. When there is no initial asphericity, the profile remains symmetric as it evolves, giving rise to spherical overdense shells propagating outward from the centre following the initial collapse. Asphericity in the initial conditions disrupts these coherent shells. In particular, increasing initial asphericity yields more complex interference patterns which begin to resemble the granularity typically of halos with more complicated formation histories.

Because granularity in the outer halo must be seeded by an explicit symmetry breaking mechanism, halos formed from the collapse of isolated overdensities such as those presented here are expected to be less granular than halos with randomly distributed neighbours and non-trivial merger histories. It is important to study halos which evolve without major merger events because these are more abundant in ULDM than in CDM due to the low-mass cutoff in the initial power spectrum \citep{Veltmaat:2018dfz}. Furthermore, high-redshift surveys will improve our ability to observe halos earlier in their evolution \citep{2021essp.confE..47D, 2017ApJ...840...92L, 2017Natur.543..397G} when fewer mergers have occurred. 
If ULDM halo characteristics are strongly dependent upon merger history, this 
would provide an additional tool to compare ULDM and CDM models.  

In our simulations, the strong oscillations in the core region can yield order-of-magnitude changes in central density on timescales that are short compared to the Hubble time, and any redshift dependence of the core-halo mass relation is thus insignificant. Moreover, the core morphology changes throughout the oscillation, and is far from solitonic at times. Hence, the core-halo relation \citep{Schive:2014hza} does not fit the majority of instantaneous halo profiles, although the match is improved by time-averaging. 

Moreover, because the peak density of the theoretical solitonic profile is proportional to the fourth power of the core mass, even a relatively small scatter around the predicted core-halo mass relation leads to a huge scatter in the predicted density profiles. This suggests that the core-halo mass relation may not be the most appropriate probe of ULDM halo characteristics, especially given its inability to account for departures from solitonic core morphologies.

Since our simulations exclude merger effects, our results may not extrapolate directly to more complex scenarios. Indeed,  simulations of ULDM halo formation through soliton collisions suggest that core oscillations in the final halo are suppressed as the number of mergers increase \citep{Zagorac:2022xic, PhysRevD.95.043519}. Isolated collapse scenarios may therefore represent extreme cases. Thus, our results emphasise the variability in ULDM halo properties and their dependence on initial conditions and formation history. Moreover, our simulations do not explore the full parameter space, so more extreme features may well be observed if the size of the initial overdensity is increased or the symmetry of the initial conditions is further reduced. 

This work raises questions about the universality of the ULDM core-halo relationship, consistent with with other analyses including \cite{Chan:2021bja, Mina:2020eik}. While the absence of a tightly constrained core-halo structure may make it difficult to identify a `smoking-gun' signature of ULDM in astrophysical observations, variability in ULDM core characteristics may be better able to accommodate the diversity of observed halos, particularly in the low-mass regime \citep{Oman:2015xda}. 

A key open question is whether the dynamical timescales of tracer stars are short enough to respond to the changes in the morphology of the core or whether stellar distributions will reflect the time-averaged dark matter profile. We plan to incorporate baryonic material into our simulations to test this explicitly and examine in detail the impact of core oscillations on stellar heating, expanding previous studies of these effects \citep{PhysRevD.101.083518, Dalal:2022rmp, Marsh:2018zyw}. The baryonic component may be initialised so as to explicitly break the symmetry in the simulations, promoting the formation of granular structure in the ULDM component and leading to further changes in the core dynamics. In particular, the gravitational coupling between ULDM and baryons may provide a dissipative mechanism for damping the oscillations of the core. If this damping is sufficiently strong, an eigenstate decomposition analysis may allow for a more precise characterisation of post-collapse dynamics \citep{Zagorac:2021qxq}, although in some cases the core oscillations may remain too large for a perturbative treatment. 

Future work will also involve simulating the collapse of several adjacent overdensities. This would be a relatively straightforward extension of the current work and would provide insight into interference effects between neighbouring halos at high resolution, which is difficult to achieve in simulations of cosmological volumes. 
In particular, it would be interesting to determine to what extent the resulting interference effects give rise to granularity in the outer halos and whether they significantly impact the core variability. It is worth noting that ``halo collapse'' and soliton formation can occur in the gravitationally fragmented inflaton condensate in the post-inflationary universe \cite{Musoke:2019ima,Niemeyer:2019aqm,Eggemeier:2020zeg,Eggemeier:2021smj,Eggemeier:2022gyo} as well as in bose stars \cite{Levkov:2018kau,Dmitriev:2023ipv}, so the dynamics investigated here will  have analogoues in other astrophysical scenarios.

Finally, while the departure from the theoretical core-halo relationship illustrated here may affect the validity of previously computed constraints on the ULDM particle mass, adaptations of the ULDM model which include multiple non-interacting ULDM fields may also weaken existing constraints \citep{Gosenca:2023yjc, Huang:2022ffc, Glennon:2023jsp}. Our work therefore motivates further analysis of the robustness of the theoretical core-halo prediction across a wider range of halo masses, formation histories, and merger scenarios, with complementary analyses also to be performed for multi-field ULDM models.

\section*{Acknowledgements}
We thank Peter Hayman, Shaun Hotchkiss,  Jens Niemeyer, Nikhil Padmanabhan, Bodo Schwabe, Y. Frank Wang  and Luna Zagorac for useful conversations. 
We acknowledge support from the Marsden Fund of the Royal Society of New Zealand and the use of New Zealand eScience Infrastructure (NeSI) high performance computing facilities, consulting support and/or training services as part of this research, supported in part by the Ministry of Business, Innovation \& Employment's Research Infrastructure programme. We acknowledge the \textsc{yt} toolkit that was used for the analysis of numerical data. 

\section*{Data Availability}

The data used in the production of this article may be shared on request to the corresponding authors.



\bibliographystyle{mnras}
\bibliography{refs} 

\clearpage
\bsp	
\label{lastpage}
\end{document}